\documentclass[conference]{IEEEtran}
\IEEEoverridecommandlockouts
\usepackage{amsmath,amssymb,amsfonts}
\usepackage{algorithmic}
\usepackage{graphicx}
\usepackage{textcomp}
\usepackage{xcolor}
\usepackage{graphicx}
\usepackage{multirow}
\usepackage{tabulary}
\usepackage{amsmath}
\usepackage{verbatim}
\usepackage{algorithm}  
\usepackage{verbatim}
\usepackage{gensymb}
\usepackage{color}
\usepackage{biblatex}
\usepackage{booktabs}
\usepackage{url}
\def\BibTeX{{\rm B\kern-.05em{\sc i\kern-.025em b}\kern-.08em
    T\kern-.1667em\lower.7ex\hbox{E}\kern-.125emX}}
\begin{document}

\title{Estimating Brain Age with Global and Local Dependencies\\
%{\footnotesize \textsuperscript{*}Note: Sub-titles are not captured in Xplore and should not be used}
\thanks{This research has been conducted using the UK Biobank Resource under Application Number 56113. 
The study is supported by grants from the Innovation Team and Talents Cultivation Program of National Administration of Traditional Chinese Medicine (NO:ZYYCXTD-C-202004), Shenzhen Longgang District Science and Technology Development Fund Project (LGKCXGZX2020002), Basic Research Foundation of Shenzhen Science and Technology Stable Support Program (GXWD20201230155427003-20200822115709001), the National Key Research and Development Program of China (2021YFC2501202), and the National Natural Science Foundation of China (62106113).
}
}

\author{\IEEEauthorblockN{Yanwu Yang\IEEEauthorrefmark{1}\IEEEauthorrefmark{2}, 
Xutao Guo\IEEEauthorrefmark{1}\IEEEauthorrefmark{2}, Zhikai Chang\IEEEauthorrefmark{1}, Chenfei Ye\IEEEauthorrefmark{1} Yang Xiang\IEEEauthorrefmark{2}, Haiyan Lv\IEEEauthorrefmark{5} and Ting Ma\IEEEauthorrefmark{1}\IEEEauthorrefmark{2}\IEEEauthorrefmark{3}\IEEEauthorrefmark{4}
}
\IEEEauthorblockA{\IEEEauthorrefmark{1}Harbin Institute of Technology at Shenzhen, China}
\IEEEauthorblockA{\IEEEauthorrefmark{2}Peng Cheng Laboratory, Shenzhen, China}
\IEEEauthorblockA{\IEEEauthorrefmark{3}Advanced Innovation Center for Human Brain Protection, Capital Medical University, Beijing, China}
\IEEEauthorblockA{\IEEEauthorrefmark{4}Xuanwu Hospital Capital Medical University, Beijing, China}
\IEEEauthorblockA{\IEEEauthorrefmark{5}MindsGo Life Science Co.Ltd, Shenzhen, China}}

%\author{\IEEEauthorblockN{Yanwu Yang, Xutao Guo, ...}
%\IEEEauthorblockA{Department of Electronic and Information Engineering, Harbin Institute of Technology at Shenzhen, China \\
%Peng Cheng Laboratory, Shenzhen, China\\
%Advanced Innovation Center for Human Brain Protection, Capital Medical University, Beijing, China\\
%National Clinical Research Center for Geriatric Disorders, Xuanwu Hospital Capital Medical University, Beijing, China\\
%MindsGo Life Science Co.Ltd, Shenzhen, China}
%}

\maketitle

\begin{abstract}

% 首先说脑龄的意义，然后说对于脑龄这种综合性生物特征，无论是特征工程还是local encoding 都容易导致较大的bias.因此我们提出了xxxx

The brain age has been proven to be a phenotype of relevance to cognitive performance and brain disease. Achieving accurate brain age prediction is an essential prerequisite for optimizing the predicted brain-age difference as a biomarker. As a comprehensive biological characteristic, the brain age is hard to be exploited accurately with models using feature engineering and local processing such as local convolution and recurrent operations that process one local neighborhood at a time. Instead, Vision Transformers learn global attentive interaction of patch tokens, introducing less inductive bias and modeling long-range dependencies. In terms of this, we proposed a novel network for learning brain age interpreting with global and local dependencies, where the corresponding representations are captured by Successive Permuted Transformer (SPT) and convolution blocks. The SPT brings computation efficiency and locates the 3D spatial information indirectly via continuously encoding 2D slices from different views. Finally, we collect a large cohort of 22645 subjects with ages ranging from 14 to 97 and our network performed the best among a series of deep learning methods, yielding a mean absolute error (MAE) of 2.855 in validation set, and 2.911 in an independent test set.

\end{abstract}


\section{Introduction}
%Aging is shown to be a significant impact on the brain structural changes, in accord with a general decline in cognitive performance and increased risk of neurodegenerative diseases such as Alzheimer’s disease \cite{abbott2011dementia} and Parkinson’s disease \cite{reeve2014ageing}.
Recently, researches have demonstrated that MRIs could be used to predict chronological age and show that the brain age, derived purely from neuroimaging data is vital to help improve detection of early-age neurodegeneration and predict age-related cognitive decline \cite{cole2017predicting}. Meanwhile, predicted age difference (PAD), the difference between predicted brain age and chronological age, correlates with the measures of mental and physical changes \cite{jonsson2019brain}. For example, positive PAD introduces that the brain is older than the actual age and the subject is experiencing accelerated aging. Furthermore, it is also shown to be associated with cognitive impairments \cite{liem2017predicting, franke2012brain}, brain injuries, and other brain diseases \cite{koutsouleris2014accelerated, kuchinad2007accelerated}. Therefore, it is an essential prerequisite to achieve accurate brain age estimating for quantifying the PAD as a biomarker.

% brain age challenging
The brain age estimation is a fine-grained recognition task, and the actual brain characteristics of the T1w image and structural changes could be hardly sensed explicitly. Recently, deep learning methods like 3D Convolution Neural Network (CNN) have been used to predict brain age and achieve promising results \cite{cole2017predicting, peng2019accurate,liu2020brain} without a prior bias or hypothesis. However, CNN methods are limited by only processing local neighborhood features and propagating signals progressively \cite{wang2018non}, where the hidden features might be lost and more inductive bias would be introduced. Recently vision Transformer models retain global image information and could relate long-range relationships between patches using self-attentions, achieving state-of-the-art performance on image classification, object detection, and semantic segmentation. The success of these models demonstrates the potential for Transformer to be used in the vision domain with the essential characteristic of encoding long-range dependencies and retaining global information. Nevertheless, studies have pointed out that long-range dependencies would also fail to work well, where a position is often less correlated far away from it, compared with those that are nearer \cite{huang2020ordnet}. Therefore, there is still a gap between encoding representations with short-range and long-range dependencies, which restricts the models' flexibility in diverse spatial scales and relationships in images \cite{huang2020ordnet}.

To address this problem, we propose a novel network for brain age estimation, called the Global and Local Dependency Network (GLDN). The GLDN sufficiently utilizes the CNNs for encoding densely-distributed local features and strengthening locality with local dependencies, and Transformer for encoding sparsely-distributed semantic concepts and establishing global dependencies. Especially, the fusion block is the basic module in our model and allows locating and aggregating the local and global concepts from CNNs and Transformer of each stage.
%In detail, each fusion block is composed of a long-range learning block, a short-range learning block, and a aggregation operation. 
Besides, we propose a new vision Transformer block, called Successive Permuted Transformer (SPT) for locating long-range dependencies of 3D medical images. The SPT leverages the spatial information of 3D images to be encoded by 2D operations by different views, which indirectly locates the spatial relationships and brings computation efficiency. Finally, we compared our model with a series of models including CNNs and vision Transformer models on a large cohort of five datasets, where the results are improved compraed with other well-estimated models.%more convincing, with MAE = 2.855, RMSE = 3.960, PCC = 0.881, SRCC = 0.871 on validation test, and MAE = 2.911, RMSE = 4.010, PCC = 0.879, SRCC = 0.869 on an independent test.

\section{Method}
\subsection{Framework}
\begin{figure}
    \centering
    \includegraphics[scale=0.54]{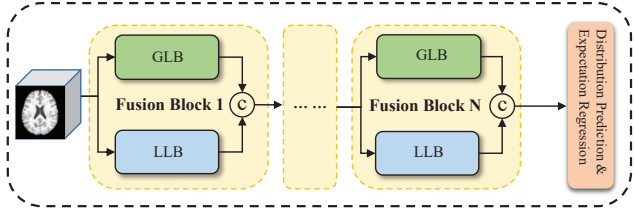}
    \caption{The framework of the GLDN network, including several fusion blocks, and a distribution prediction and expectation regression module. Each fusion block(in yellow) is composed of a global learning block (GLB, in green), a local learning block (LLB, in blue), and an aggregation operation (shown as a symbol C).
    }\label{whole}
\end{figure}

We illustrate the sketch map of the GLDN network in Fig.~\ref{whole}, which is composed of several fusion blocks, and a classifier with a label distribution prediction and expectation regression module.
Each fusion block is built with a global learning block, a local learning block, and an aggregation operation.

\begin{figure*}
    \centering
    \includegraphics[scale=0.7]{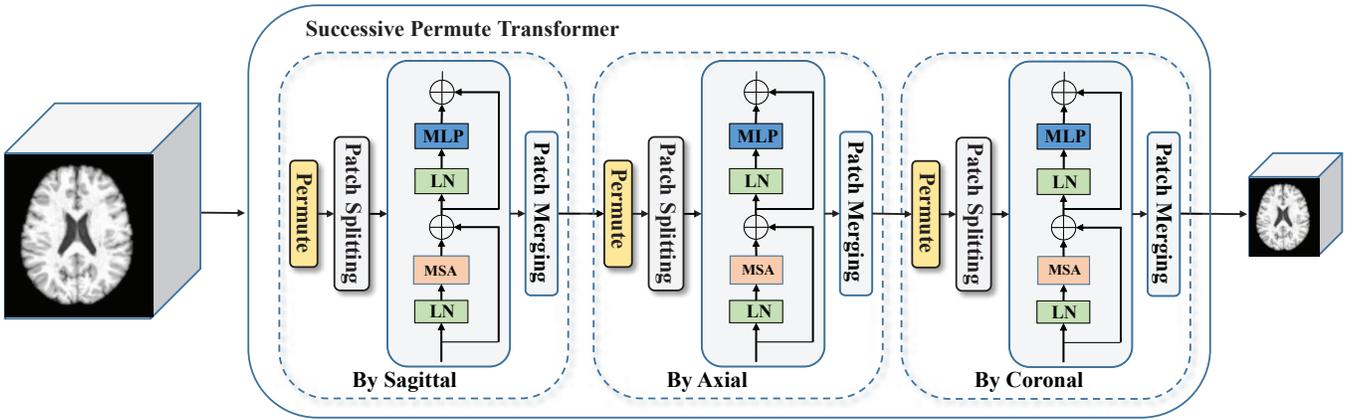}
    \caption{The design of a SPT block, where the 3D information is learned by three successive Transformer. Permutation operations leverages the framework to relate spatial context from one view at one time.}
    \label{spt}
\end{figure*}

\subsubsection{\textbf{Local Learning Block}}
In detail, two CNN blocks are embedded in the local learning block. Each CNN block consists of a 3D convolution layer with a kernel size of 3, padding size of 1, and batch normalization, a ReLU activation, and a max-pooling layer with a pooling size of 2. With two CNN blocks, the feature size of input would be reduced into $(\frac{1}{4}, \frac{1}{4}, \frac{1}{4})$. The local Learning Block could be denoted as:
\begin{equation}
    x_{local}^l=LLB(x^l) = CB(CB(x^l))
\end{equation}
\begin{equation}
    CB(x^l) = MaxPool(BN(ReLU(Conv(x^l))))
\end{equation}
where $x^l$ is the input of the $l-th$ fusion block, $x_{local}^l$ is the output of a local learning block, $LLB$ denotes the local learning block, and $CB$ denotes a CNN block.
%Each encoding block is composed of a SPT block and two CNN blocks, where the CNN and SPT learn densely-distributed local patterns and sparsely-distributed, nonlocal semantic concepts respectively. At the end of each encoding block, the features are combined using a concatenation operation.

% SPT
\subsubsection{\textbf{Global Learning Block}}
% 这里还得加点描述，说为什么这么设计是更好的。原因在于传统的T参数量过大，直接对3D建模导致层数无法特别深，导致学习能力下降，收敛慢。
The global learning block is built with successive permute Transformer blocks to capture global information with full-range dependencies. Apart from the original ViT structure, the SPT, shown in Fig.~\ref{spt}, is designed to better suit 3D medical images with three sequential Transformer parts, where each part locates relationships of slices separated from different views (Sagittal, Axial, Coronal). In detail, each part consists of a permute operation, a patch splitting layer, Transformer layers, and a patch merging layer. 

To reduce numerous parameters in modeling 3D medical images using Transfromer, the 3D input is permuted along each axis and cropped into 2D slices. For example, a permute operation would transform an input in the size of $96 \times 114 \times 96$ into 96 slices with each size of $114 \times 96$ along the first axis. The slice would be continued to be segmented into non-overlapping patches by patch splitting.

% 由于3D医疗图像的维度高，直接使用t进行特征编码会导致模型的超参数过多，最终导致训练困难，难以收敛

The Transformer encoders consist of multi-head self-attention blocks (MSA), layer norms, and fully connected feed-forward blocks. The input tokens $x_t \in R^{N \times d}$ are linear projected into $qkv$ spaces, where queries (Q), keys (K), values (V), and the output, a weighted sum of the values are computed as
\begin{equation}
    Attention(Q,K,V) =  Softmax(\frac{QK^T}{\sqrt{d_k}})V
\end{equation}
Finally, the multi-head attention layer is defined by combining multiple attentions. The outputs of several self-attention blocks are concatenated and sent into the patch merging layers. Through computing dot-product, the similarity between different tokens is calculated, resulting in long-range and global attention.

The patch merging layers are implemented for feature map compression via concatenating the features of each group of 2 $\times$ 2 neighboring patches and linear projection, at the same time producing hierarchical representations, by reference to the design in \cite{liu2021swin}. The patch merging layer is implemented followed by the Transformer layer. Within a SPT, the input would be reduced into $(\frac{1}{4} \times \frac{1}{4} \times \frac{1}{4})$ with each patch merging layer downsampling the feature size of one view into $\frac{1}{2} \times \frac{1}{2}$. The design of successive Transformer indirectly realizes the spatial position relationship learning through permute operations. 
Finally, the proposed GLDN network can be noted as:
\begin{equation}
    x_{local}^l = LLB(x^l)
\end{equation}
\begin{equation}
    x_{global}^l = GLB(x^l)
\end{equation}
\begin{equation}
    x_{fusion}^l = Aggregate(\{x_{local}^l, x_{global}^l\})    
\end{equation}

where, $l$ denotes the $l$-th layer, and $GLB$ denotes the global learning block.

\begin{comment}

Finally, the proposed GLDN network can be noted as Algorithm.\ref{alg:A}:
\begin{algorithm}
    \caption{GLDN Network}
    \label{alg:A}
    \begin{algorithmic}
        \STATE {$X :$ the input; $X_p$: the tokenized patch input.}\
        \STATE {$l :$ fusion stage number.}\
        \STATE {$\alpha :$ the view of 3D input $\{0,1,2\}$ represents \{Sagittal, Axial, Coronal\}.}\
        \STATE {$y: $ the output.}\
        \STATE {$LLB: local learning block$}\
        \STATE Initialize $X = input$;\
        \FOR{$l = 1 \in [1, N]$}
            \STATE $\hat{X} = X$\
            \FOR{$\alpha \in \{0, 1, 2\}$}
                \STATE $X^{l}_{p,\alpha} = Split(Permute(X, \alpha))$;\
                \STATE $\hat{X}^{l}_{p,\alpha} = MSA(LN(X^{l}_{p,\alpha})) + X^{l}_{p,\alpha}$;\
                \STATE $\Bar{X}^{l}_{p,\alpha} = MLP(LN(X^{l}_{p,\alpha})) + X^{l}_{p,\alpha}$;\
                \STATE $\hat{X} = Merge(\Bar{X}^{l}_{p,\alpha}$);\
            \ENDFOR
            \STATE $\hat{X} = Aggregate(\hat{X},LLB(X^{\prime}));$
        \ENDFOR
        \STATE $y = Softmax(FC(GAP(\hat{X})))$;\
    \end{algorithmic}
\end{algorithm}
\end{comment}

% 这种连续层的设计通过permute间接的实现了对三维医学图像的空间位置关系学习

\subsection{LDL \& expectation regression and loss function} 
%Naturally, the brain age is difficult to be estimated with an explicit value and easier to be described in a way like "around 40". Such label ambiguity always happens and could not be described with the image or even voxels, especially for brain age. In terms of this, the LDL is proposed to transform a single value age into a label distribution, where the regression problem task is transformed into a label distribution learning problem. This strategy forces the deep learning regression model to take care of the ambiguity among labels \cite{gao2017deep} and achieves promising results in facial age \cite{geng2014facial,yang2015deep} and brain age estimation \cite{hu2019deep,liao2020multi}.

%However, the performance is affected by the label distribution, where the hyper-parameter $\theta$ decides the range of distribution. Furthermore, such indirect regression would maximize the similarity between the source and target label distributions, however little improvement in label prediction. To overcome this limit, 
In this paper, we leverage the label distribution learning combined with the expectation regression as the loss function. This strategy forces the deep learning regression model to take care of the ambiguity among labels \cite{gao2017deep}. In this paper, all the ages ($y \in R$) of healthy subjects range from 14 to 97. And we define the label set as $L = (l_k| k = 14, 15, ..., 97)$, and $\Delta l = 1$ as the discrete step size. The probability density function of normal distribution is chosen to generate the ground-truth $(q_k|k = 14, 15, ..., 97)$ with a hyper-parameter $\theta$:
\begin{equation}
    y = \sum_{k}{l_kq_k}, q_k \in (0,1)
\end{equation}
\begin{equation}
    q_k = \frac{p_k}{\sum_{k}{p_k}}
\end{equation}
\begin{equation}
    p_k = \frac{1}{\sqrt{2\pi}\theta}e^{-\frac{(l_k - y)^2}{2\theta^2}}
\end{equation}

\begin{comment}
\begin{figure}
    \centering
    \includegraphics[scale=0.53]{loss.eps}
    \caption{The label distribution prediction and expectation regression module. We firstly generate a label distribution and then jointly optimize label distribution learning and expectation regression.}
    \label{loss}
\end{figure}

\begin{figure*}
    \centering
    \includegraphics[scale=0.7]{heat2.eps}
    \caption{Example of feature visualization of SPT and CNN blocks, where the features of ten subjects are shown in the 2D form. The example features are extracted by the first learning stage and over-plotted the values with a threshold of 95\%. Different regions of interest learned by SPT and CNNs are shown in the second row and the third row respectively, where the brighter regions introduce more attention. It shows that the Transformer is pruning to locate sparsely-distributed, higher-order semantic concepts and convolution encodes densely-distributed local patterns.}
    \label{heat2}
\end{figure*}
\end{comment}

%Finally, the classifier with LDL and expectation regression is built with two 3D-CNN blocks with a kernel size of [3, 1], 448 and 64 features maps respectively and a global average pool (GAP), and linear projection into brain age representation.. Each 3D-CNN block is implemented with batch normalization and a ReLU activation. Finally, the GAP layer and linear projection synthesize mixture high-level features for brain age prediction into 84.

%\subsection{Loss Functions}
The goal of the label distribution prediction is to maximize the similarity between $q_k$ and the predicted distribution $\hat{q_k}$. The Kullback-Leibler divergence is employed as the measurement of the dissimilarity between ground-truth label distribution and prediction distribution:
\begin{equation}
    L_{kl} = \sum_{k}{q_klog{\frac{q_k}{\hat{q_k}}}}
\end{equation}
And the expectation regression takes the predicted distribution and the label set $L$ as inputs. The expectation regression module minimizes the error between the expected value $\hat{y}$ and ground-truth $y$. The $L_1$ loss is used as the error measurement:
\begin{equation}
    L_{mae} = |\hat{y} - y| = |\sum_{k}{\hat{q_k}l_k} - y|
\end{equation}
Totally, the weighted combination of $L_{kl}$ and $L_{mae}$ with the weight $\lambda$ is employed as our loss function:

\begin{equation}\label{equ:combine_loss}
    L = L_{kl} + \lambda{L_{mae}} = \sum_{k}{q_klog\frac{q_k}{\hat{q_k}}} + \lambda|\sum_{k}{\hat{q_k}l_k} - y|
\end{equation}
\section{Experiments}
\subsection{Datasets}
The methods were evaluated on T1-weighted MR images from a large cohort consisting of IXI database (\url{http://brain-development.org}), the Alzheimer’s Disease Neuroimaging Initiative (ADNI) \cite{jack2008alzheimer}, UK Biobank\cite{sudlow2015uk,miller2016multimodal}, the Open Access Series of Imaging Studies (OASIS)\cite{marcus2010open}, and 1000 Functional Connectomes Project (1000-FCP, \url{http://www.nitrc.org/projects/fcon_1000}). Only healthy subjects were selected in our experiments. A total of 22645 T1-weighted MRI images of subjects aging between 14 and 97 years old are selected to form our cohort. All data were acquired at either 1.5T or 3T T1-weighed MRI.

\subsection{Data Preprocessing}
%All the data were processed following the pipeline, which is described in the online brain imaging documentation (UK Biobank Brain Imaging Documentation). Notably, the data from UK Biobank were already preprocessed\cite{alfaro2018image}. 
All data were all processed with the same pipeline, including image FOV truncation \cite{jenkinson2012fsl}, AC-PC align, brain skull stripping, bias field correction \cite{sled1998nonparametric}, and linear-registration into the standard MNI space. Additionally, z-score normalization is employed to narrow the gap between different data centers. It is proved to improve the synthesis results and is vital for successful deep learning-based MR image synthesis \cite{reinhold2019evaluating}. After preprocessing, all images are down-sampled by linear-registering into the standard 2$mm^3$ MNI space and padded into the size of $96 \times 112 \times 96$ for successive non-overlapping downsampling.

%##################################################
\begin{table*}
\begin{center} 
\setlength{\tabcolsep}{3.5mm}
\footnotesize
\caption{Performance of brain estimations using different models.}\label{tab2}
\begin{tabular}{c|c|cccc|cccc}
\hline
\specialrule{0em}{1pt}{1pt}
& & \multicolumn{4}{c}{Validation Set} & \multicolumn{4}{|c}{Independent Test Set}\\
\hline
Type & Models & MAE &RMSE & PCC & SRCC & MAE &RMSE & PCC & SRCC \\[1.5pt]
\hline
Machine learning & XGBoost+LightGBM  & 4.290 & 7.130 & 0.815 & 0.818 & 4.298 & 7.150 & 0.808 & 0.814\\[1.5pt]
\hline
\multirow{5}*{\shortstack{CNNs}}& Resnet18 &3.265 & 4.386   &0.871   &0.867 & 3.383 & 4.612 & 0.858 & 0.854\\[1.5pt]
&Resnet34 &3.204 & 4.314   &0.873   &0.867 & 3.334 & 4.640 & 0.861 & 0.851\\[1.5pt]
&Resnet50 &3.226 & 4.553   &0.871   &0.867 & 3.349 & 4.817 & 0.854 & 0.850\\[1.5pt]
&SFCN &2.993 & 4.097   &0.875   &0.869 & 3.093 & 4.228 & 0.872 & 0.856\\[1.5pt]
&TSAN (first-stage) &2.948 & 4.150   &0.874   &0.868 & 3.076 & 4.350 & 0.861 & 0.866\\[1.5pt]
\hline
%RNNs & RNN in \cite{} & \\
\multirow{7}*{\shortstack{ \\ \\Transformer \\Models}} 
&ViT & 3.419 & 4.613 & 0.863 & 0.865 & 3.536 & 4.859 & 0.848 & 0.851 \\[1.5pt]
&DeTR-1 & 3.335 & 4.509 & 0.866 & 0.865 & 3.344 & 4.517 & 0.863 & 0.854 \\[1.5pt]
&DeTR-2 & 2.920 & 4.101 & 0.876 & 0.866 & 2.997 & 4.260 & 0.863 & 0.854\\[2pt]
&DeTR-3 & 2.933 & 4.122 & 0.876 & 0.868 & 3.047 & 4.286 & 0.867 & 0.855 \\[1.5pt]
& Nested Transformer & 3.112 & 4.392 & 0.873 & 0.866 & 3.234 & 4.630 &  0.844 & 0.850\\[1.5pt]
\cline{2-10}
%\specialrule{0em}{1pt}{1pt}
& Ours (w/o CNN)& 3.041 & 4.153 & 0.873 & 0.868 & 3.167 & 4.335 & 0.866 & 0.854 \\[1.5pt]
&Ours & \textbf{2.855} & \textbf{3.960} & \textbf{0.881} & \textbf{0.871} & \textbf{2.911} & \textbf{4.010} & \textbf{0.879} & \textbf{0.869}\\[2pt]
\hline
\end{tabular}
\end{center}
\end{table*}
%##################################################

\subsection{Experimental Setting}
%test set
For comparison, we divided all the data samples into three subsets including training set (80\%), validation set (10\%), and a test set (10\%). The test set is fixed and a cross-validation with 4 fold was performed on the rest samples. The performance is evaluated by the mean absolute error (MAE), root mean squared error (RMSE), Pearson correlation coefficient (PCC), and Spearman's rank correlation coefficient (SRCC).
%An independent test dataset including 2265 samples (10\%) was randomly selected to test the model with stratified sampling by grouping every 10 ages with relatively uniform probability.
%validation set
%Stratified cross-validation with 4 fold was performed on the rest samples. This could reduce the age bias of our model and improve the generalization ability to datasets with different age distributions.

Our proposed GLDN is embedded with two fusion blocks to ensure the integrity of slicing and downsampling. The first fusion block is built with an SPT block with a patch size of $[8 \times 8]$, and two CNN blocks with the channel number of [16, 32]. The second fusion block receives the concatenation of the first fusion block with a channel of 40. Within the second fusion block, the patch size of the SPT is set as $[2 \times 2]$, and the convolution channel is set as [64, 128].

For better comparison, machine learning methods with feature engineering and deep learning methods with CNNs, and Transformer networks are both carried out. An ensemble model of XGBoost \cite{chen2015xgboost} and LightGBM \cite{ke2017lightgbm} methods was carried out, using relative volume fraction of the brain regions segmented by FastSurfer \cite{henschel2020fastsurfer}.
A series of end-to-end CNN based methods including 3D-ResNet, SFCN \cite{peng2019accurate}, and TSAN \cite{liu2020brain} are trained and implemented. For ResNet, the original 2D operations were replaced by 3D and a dropout with drop rate = 0.5 was applied before the final fully connected layer. Here we implement the first stage of TSAN for fair comparison.
Besides, Transformer methods like ViT, DETR \cite{carion2020end}, Nested Transformer are also compared. The depth of layers and number of the attention heads are tuned according to different architectures with a grid search of [4, 6, 8, 12]. These models receive 2D images as input for default and are modified to suit 3D medical images by replacing 2D operations into 3D.

In addition, we implement the DeTR design with different numbers of layers, where 1/2/3 CNN blocks are compared in DeTR-1/2/3 respectively. Especially, the nested Transformer is modified by reference to the \cite{han2021transformer} and is carried out using an inner Transformer (depth: 6, heads: 8) encoding 2D slices, and an outer Transformer encoding spatial information (depth: 8, heads: 12).

All the models were trained from scratch with a initialized learning rate of 1e-4, and a batch size of 128. The learning rate is increased to 1e-4 in 200 warmup epochs. The $\lambda$ was initialized with 0 and set to 1 when the validation loss had not been decreased for 50 epochs. Models were trained using a stable adaptive optimizer, Adam \cite{kingma2014adam}, with a L2 weight decay coefficient = 0.00005, $\beta_1 = 0.9$, and $\beta_2 = 0.999$.
The best model was obtained based on the validation loss and an early stopping criterion was imposed when the validation loss did not improve for 80 epochs. To reduce the risk of overfitting, two data argumentation methods were applied during training, consisting of random rotation and random image shifting. The rotation angles were between $-10^{\degree}$ and $10^{\degree}$ and the input was random shifted by between -5 and 5 voxels along every axis with equal probability. All the experiments are carried out using Pytorch on 8 NVIDIA-Tesla V100 GPU devices.

\begin{comment}

\begin{figure*}
    \centering
    \includegraphics[scale=0.5]{boxplot.eps}
    \caption{A boxplot of the performance on different data centers of all the models, including a machine learning method (the ensemble of XGBoost and LightGBM, shown as ML), DeTR-1/2/3, ResNet-18/34/50, SFCN, TSAN (first-stage), ViT, Nested Transformer, GLDN (with and without CNNs), where all the results are evaluated on the test set. The median line and the mean line are shown in yellow and green. And the best result is compared using a dotted line.}
    \label{boxplot}
\end{figure*}
\end{comment}
\section{Results}
The detailed evaluation results are shown in Table.~\ref{tab2}. The machine learning method using feature engineering with an ensemble of XGBoost and LightGBM did not perform as well as deep learning methods. With the layers getting deep in ResNet, the performance increases first and remains or even decreases a bit, where the ResNet-34 achieves the best among all the ResNet families, with the MAE of 3.204, RMSE of 4.314, PCC of 0.873, and SRCC of 0.869 in the validation set, and the MAE of 3.334, RMSE of 4.640, PCC of 0.861, and SRCC of 0.851 in the test set. The design of light fully convolution model with large number of channels and dense model with asymmetric convolution achieve promising and comparable results in CNN models, better than most Transformer based models. 
Transformer based model also achieves promising results, where the DeTR with 2 CNN blocks performs the best with the MAE of 2.920, RMSE of 4.101, PCC of 0.876, and SRCC of 0.868 in the validation set, and the MAE of 2.997, RMSE of 4.260, PCC of 0.863, and SRCC of 0.854 in the test set. With the number of CNN blocks increasing, the performance reached a plateau with excessive abstract low-level features. 

We show ablation experiments on our proposed GLDN network without CNN stressing on locality. Although we excluded the CNN for discarding the localized dependencies, our design of SPT is better than methods with pure vision Transformers (ViT, Nested Transformer) for 3D medical image learning with continuously encoding relationship along different axes and achieves the MAE of 3.041, RMSE of 4.153, PCC of 0.881, and SRCC of 0.871 in the validation set, and the MAE of 3.167, RMSE of 4.335, PCC of 0.866, and SRCC of 0.854 in the test set. 
Compared with CNN models and Transformer models, the methods (DeTR, GLDN) of using the CNN and Transformer together for encoding features achieve the best. 
Finally, our proposed GLDN generally obtains the best results with the lowest MAE and the highest PCC (MAE: 2.855, RMSE: 3.960, PCC: 0.881, SRCC: 0.871 in the validation set, and MAE: 2.911, RMSE: 4.010, PCC: 0.879, SRCC: 0.869 in the test set). 

\section{Conclusion}
In this paper, the GLDN was proposed to predict individual brain age based on brain MRI images with Transformer encoding global representations and establishing global dependencies and CNN stressing on locality with local dependencies. The architecture improves feature diversity and aggregates the multi-scale information. 
In our experiments, the combination of convolutions and Transformer would achieve promising results among all the models, where our proposed model achieves the optimal, yielding an MAE of 2.911, RMSE of 4.010, PCC of 0.879, and SRCC of 0.869 on the independent test set. Overall, we suspect that the coordination between Transformer and convolution has a great potential for analyzing neuroimaging-based individualized prediction of the clinical or behavioral phenotype.

%\section*{Acknowledgment}
%This research has been conducted using the UK Biobank Resource under Application Number 56113. This study is supported by grants from the National Key Research and Development Program of China (2018YFC1312000), Basic Research Foundation of Shenzhen Science and Technology Stable Support Program (GXWD20201230155427003-20200822115709001), China Postdoctoral Science Foundation funded project (2021M691686), and the National Natural Science Foundation of China (62106113).

\printbibliography

\end{document}